\documentclass[twocolumn,prl,aps,epsfig]{revtex4}
\usepackage{graphicx}
\tighten
\newcommand{\beq}{\begin{eqnarray}}          
\newcommand{\eeq}{\end{eqnarray}}

\begin{document}
\draft
%\twocolumn
%\preprint{dvi file made on \today}
\input epsf.sty

\title
{Breakdown of One-Paramater Scaling in Quantum Critical Scenarios for the High-Temperature Copper-oxide Superconductors}
\author{ Philip Phillips}
\affiliation{Loomis Laboratory of Physics,
University of Illinois at Urbana-Champaign,
1110 W.Green St., Urbana, IL., 61801-3080}
\author{Claudio Chamon}
\affiliation{Department of Physics, Boston University, Boston, MA. 02215}
\begin{abstract}
 We show that if the excitations which become gapless at a quantum critical
  point also carry the electrical current, then a  resistivity linear in temperature, as is observed in the copper-oxide high-temperature superconductors,
  obtains only if the dynamical exponent, $z$, satisfies the unphysical
  constraint, $z<0$.   At fault here is the universal scaling hypothesis that,
  at a continuous phase transition, the only relevant length scale is the
  correlation length. Consequently, either the electrical current in the
  normal state of the cuprates is carried by degrees of freedom which do not
  undergo a quantum phase transition, or quantum critical scenarios must
  forgo this basic scaling hypothesis and demand 
that more than a single correlation length scale
  is necessary to model transport in the cuprates.
\end{abstract}

\maketitle

\begin{figure}[t]
\centering
\includegraphics[width=7cm]{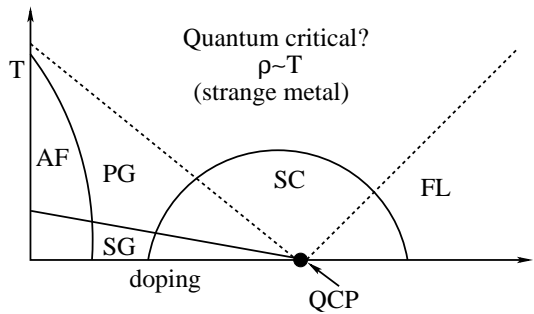}
\caption{Heuristic phase diagram of the high temperature copper oxide
  superconductors as a function of temperature and hole concentration
  (doping). The phases are as follows: AF represents antiferromagnet, SG, the
  spin glass and SC the superconductor. The spin-glass phase terminates at a
  critical doping level (quantum critical point, QCP) inside the dome.  The
  dashed lines indicate crossovers not critical behaviour. In this context PG
  and FL represent the pseudogap and Fermi liquid phases in which
  respectively the single-particle spectrum develops a dip and the transport
  properties become more conventional. The strange-metal behaviour,
  $T-$linear resistivity, in the funnel-shaped regime has been attributed to
  quantum critical behaviour.  A scaling analysis of the conductivity at the
  quantum critical point rules out this scenario, however.}
\label{fig1}
\end{figure}

The central problem posed by the normal state of the high-temperature copper
oxide superconductors (cuprates) is the riddle of the $T-$linear
resistivity\cite{batlogg}.  Namely, over a funnel-shaped region in the
temperature-doping plane (as in Fig. \ref{fig1}), the resistivity is a linear function of temperature
rather than the $T^2$ dependence indicative of typical metals.  Equally
perplexing is the persistence of this transport anomaly to unusually high
temperatures, roughly $1000K$.  At present, there is no consensus as to the
origin of this robust phenomenon.  However, two scenarios, 1) marginal fermi
liquid (MFL) phenomenology\cite{mfl} and 2) quantum criticality\cite{qc1,qc2}
have been advanced.  The former rests on the empirical observation\cite{mfl}
that the self energy needed to describe the broad line shapes in angularly
resolved photoemission (ARPES) also yields a scattering rate, and hence a
resistivity, that scales linearly with temperature.  In contrast, quantum
criticality provides a first-principles account.  At the quantum critical
coupling, or in the quantum critical regime, the only energy scale governing
collisions between quasiparticle excitations of the order parameter is
$k_BT$.  Consequently, the transport relaxation rate scales as
\begin{eqnarray}
\label{lint} 
\frac{1}{\tau_{\rm tr}}\propto\frac{k_BT}{\hbar}, 
\end{eqnarray} 
thereby implying a $T-$linear resistivity if (naively) the scattering rate is
what solely dictates the transport coefficients.  While MFL fitting\cite{mfl}
also achieves a scattering rate of this form, a $T=0$ phase transition is not
necessarily the operative cause.  The fact that temperature emerges as the
only scale in the quantum critical regime regardless of the nature of the
quasiparticle interactions is a consequence of universality. Eq. (\ref{lint})
is expected to hold as long as the inequalities $T>|\Delta|$ and
$t<1/|\Delta|$ are maintained, $\Delta$ the energy scale measuring the
distance from the critical point and $t$ the observation time. The energy scale $\Delta\propto \delta^{z\nu}$ varies as a
function of some tuning parameter $\delta=g-g_c$, where $\nu$ is the
correlation length exponent and $z$ is the dynamical exponent. At the
critical coupling $\delta=0$ or $g=g_c$, the energy scale $\Delta$ vanishes. Ultimately, the
observation time constraint, $t<1/|\Delta|$ implies that only at the quantum
critical point does the $T-$linear scattering rate obtain for all times. That
the quantum critical regime is funnel-shaped follows from the inequality
$T>|\Delta|$.
The funnel-shaped critical region should be bounded by a temperature $T_{\rm
  upper}$ above which the system is controlled by high-energy processes. That
quantum criticality is operative up to temperatures of order $T=1000K$ in the
cuprates is questionable, but we relax this criticism in carrying out our
analysis of the scaling of the resistivity.

Because Eq. (\ref{lint}) is valid for any $T=0$ phase transition, it has been
quickly adopted as the explanation of choice for the $T-$linear resistivity
in the cuprates.  In fact, there has been no
paucity\cite{ando1,boeb,valla,christos,marel} of candidate quantum
critical points proposed for the cuprates: 1) at $1/8^{\rm th}$-hole doping
in Bi$_2$Sr$_{2-x}$La$_x$CuO$_{6+\delta}$ at 58T, a transition\cite{ando1}
occurs between an anisotropic 2D and a 3D superconductor, 2) near optimal
doping, the Hall coefficient\cite{boeb} exhibits a significant break,
indicating a fundamental restructuring of the Fermi surface, and 3) in
La$_{2-x}$Sr$_x$Cu$_{0.95}$Zn$_{0.05}$O$_4$ a spin glass state
terminates\cite{christos} at roughly optimal doping, $x\approx 0.19$.
Regardless of its origin, a $T=0$ phase transition near optimal doping can, in principle,
account for the funnel-shaped $T-$linear region seen in early transport
experiments.  However, several experiments\cite{raffy,ando2} call into
question the very existence of such a wide region.  Namely, Raffy, et. al.\cite{raffy} and
also Ando, et.
al.\cite{ando2} find that the $T-$linear region is not a region at all,
existing only at optimal doping. While Ando, et. al.\cite{ando2} argue that
the absence of a triangular region (in the $T-x$ plane) near optimal doping
strongly suggests that quantum criticality is not the cause of the $T-$linear
resistivity, an equally valid explanation is that the time constraint
$t<1/|\Delta|$ is violated except at optimal doping.  Nonetheless, optical
transport measurements\cite{marel} find that contrary to theoretical
predictions\cite{ds}, the optical conductivity does not obey the predicted
scaling law $T^{-\mu} f(\omega/T)$ with a constant $\mu$ as $\omega/T$ is
varied.  In contrast, they find\cite{marel} that $\mu=1$ for $\omega/T<1.5$
and $\mu\approx 0.5$ for $\omega/T>3$. In addition, in the quantum critical
regime of other strongly correlated systems, such as the
heavy-fermions\cite{stewart}, the resistivity exhibits a non-universal
algebraic temperature dependence of the form $\rho\propto\rho_0+AT^\alpha$
with $0.3<\alpha<2.0$ and $A<0$ or $A>0$.

Motivated by such experiments, we examine what conditions must hold for
$T-$linear resistivity to be compatible with the universally accepted
assumption that at a continuous quantum critical point, the only relevant
length scale is the correlation length. We obtain, using the single scale
hypothesis and the fact that electric charge is conserved, a very general
scaling law for the electric conductivity near a quantum critical point. This
scaling law must hold irrespective of the microscopic details of the theory,
and regardless of the quantum statistics of the charge carriers, be they
bosons or fermions. From the scaling law, we find that $T-$linear resistivity
obtains only if the dynamical exponent $z<0$, which is an unphysical negative
value. Consequently, no consistent account of $T-$linear resistivity is
possible if the quantum critical modes carry the electrical charge. We
conclude that either the degrees of freedom that are responsible for the
$T-$linear resistivity in the cuprates are not undergoing a quantum phase
transition, or that quantum critical scenarios must reliquish the simple
single scale hypothesis to explain the resistivity law in the cuprates.

To proceed, we derive a general scaling form for the conductivity near a
quantum critical point. Consider a general action $S$, the microsocopic
details of which are unimportant. An externally applied electromagnetic
vector potential $A^\mu$, $\mu=0,1,\dots,d$, couples to the electrical
current, $j_\mu$, so that
\begin{equation} 
S\to S+\int d\tau \;d^dx\; A^\mu\;j_\mu.
\end{equation} 
The one-parameter scaling hypothesis in the context of quantum systems is that spatial
correlations in a volume smaller than the correlation volume, $\xi^d$, and
temporal correlations on a time scale shorter than $\xi_t\propto\xi^z$ are
small, and space-time regions of size $\xi^d\xi_t$ behave as independent
blocks. With this hypothesis in mind, we write the scaling form for the
singular part of the logarithm of the partition function by counting the
number of correlated volumes in the whole system:
%\begin{equation} 
%\ln Z=\frac{L^d\beta}{\xi^d\xi_t}F(\delta\xi^{d_\delta}, A^i_\omega
%f(\omega\xi_t)\xi^{d_A})
%\;,
%\label{eq:lnZ}
%\end{equation} 
\begin{equation} 
\ln Z=\frac{L^d\beta}{\xi^d\xi_t}\;F(\delta\xi^{d_\delta}, 
\{A^i_{\lambda}\; \xi^{d_A}\})
\;,
\label{eq:lnZ}
\end{equation} 
In this expression, $L$ is the system size, $\beta=1/k_BT$ the inverse
temperature, $\delta$ the distance from the critical point, and $d_\delta$
and $d_A$ the scaling dimensions of the critical coupling and vector
potential, respectively. The variables $A^i_{\lambda}=A^i(\omega=\lambda
\xi_t^{-1})$ correspond to the (uniform, $k=0$) electromagnetic vector
potential at the scaled frequency $\lambda=\omega\xi_t$, and $i=1,\dots,d$
labels the spatial components. Two derivatives of the logarithm of the
partition function with respect to the electromagnetic gauge $A^i(\omega)$,
\begin{eqnarray}
\sigma_{ij}(\omega,T)&=&
\frac{1}{L^d\beta}\;\frac{1}{\omega}\frac{\delta^2\ln Z}
{\delta A^i({-\omega}) \delta A^j({\omega})} 
\nonumber\\
&=& \xi^{-d}\;\frac{\xi_t^{-1}}{\omega}
\;
\;\xi^{2d_A}\;\frac{\delta^2}{\delta A^i_{-\bar\lambda}\delta A^j_{\bar\lambda}}
F(\delta=0,\{A^i_\lambda=0\})\Big|_{\bar\lambda=\omega\xi_t}
\nonumber\\
&=&\frac{Q^2}{\hbar}\;\xi^{2d_A-d}\;\Sigma_{ij}(\omega\xi_t), 
\end{eqnarray} 
determine the conductivity for carriers with charge $Q$. We have explicitly
set $\delta=0$ as our focus is the quantum critical regime. At finite
temperature, the time correlation length is cutoff by the temperature as
$\xi_t\propto 1/T$, and $\xi_t\propto \xi^z$. The engineering dimension of
the electromagnetic gauge is unity $(d_A=1)$. Charge conservation prevents
the current operators from acquiring an anomalous dimension; hence, that
$d_A=1$ is exact~\cite{Wen1992}. We then arrive at the general scaling form
\begin{equation}
\label{genscal}
\sigma(\omega,T)=\frac{Q^2}{\hbar}\;T^{(d-2)/z}
\;\Sigma\left(\frac{\hbar\omega}{k_B
    T}\right) 
\end{equation} 
for the conductivity where $\Sigma$ is an explicit function only of the
ratio, $\omega/T$. (We have dropped the $ij$ tensor indices for simplicity.)
This scaling form generalizes to finite $T$ and $\omega$ the $T=0$ frequency
dependent critical conductivity originally obtained by Wen~\cite{Wen1992}.
The generic scaling form, Eq.~(\ref{genscal}), is also in agreement with that
proposed by Damle and Sachdev\cite{ds} in their extensive study of
collision-dominated transport near a quantum critical point (see also the
scaling analysis in Ref.~\cite{book1}). What the current derivation lays
plain is that regardless of the underlying statistics or microscopic details
of the Hamiltonian, be it bosonic (as in the work of Damle and
Sachdev\cite{ds}) or otherwise, be it disordered or not, the general scaling
form of the conductivity is the same. A simple example where such scaling
formula for the conductivity applies is the Anderson metal-insulator
transition in $d=2+\epsilon$, which can be thought of as a quantum phase
transition where the dimensionless disorder strength is the control
parameter~\cite{Wegner,Gang4}.

In the dc limit,
\begin{eqnarray}\label{dclimit}
\sigma(\omega=0)=\frac{Q^2}{\hbar}\;\Sigma(0)\;\left(\frac{k_BT}{\hbar
    c}\right)^{(d-2)/z} .
\end{eqnarray} 
In general $\Sigma(0)\ne 0$\cite{finite}. Else, the conductivity is determined entirely by the non-singular and hence non-critical part of the free energy.
The cuprates are anisotropic 3-dimensional systems.  Hence, the relevant dimension for the critical modes is $d=3$ not $d=2$. In the latter case, the temperature prefactor is constant. For $d=3$, we find that $T-$ linear resistivity obtains
only if $z=-1$.  Such a negative value of $z$ is unphysical as it implies
that energy scales diverge for long wavelength fluctuations at the critical point.   In fact, that
the exponent of the temperature prefactor in Eq. (\ref{genscal}) is strictly
positive is inconsistent with the Drude formula for the conductivity.
Consider the work of van der Marel, et.  al.\cite{marel} in which a Drude
form for the conductivity,
\begin{eqnarray}
\label{drude}
\sigma_{\rm Drude}=\frac{1}{4\pi}\frac{\omega^2_{\rm pl}\tau_{\rm
    tr}}{1+\omega^2\tau^2_{\rm tr}}, 
\end{eqnarray} 
was used to collapse their optical conductivity to a function of $\omega/T$
($\omega_{\rm pl}$ is the plasma frequency).  Because $\tau_{\rm tr}\propto
1/T$, the Drude form for the conductivity is consistent with the critical
scaling form for the conductivity, Eq.~(\ref{genscal}), only if $z=-1$. The
presence of another energy scale\cite{enscal} in the Drude formula, namely the plasma
frequency, is also at odds with the scaling form in Eq.~(\ref{genscal}).  On
dimensional grounds, the $z=-1$ result in the context of the Drude formula is
a consequence of compensating the square power of the plasma frequency with
powers of the temperature so that the scaling form Eq.~(\ref{genscal})) is
maintained. Hence, data collapse according to the Drude formula is not an
indication that the universality which underlies the scaling form of Eq.
(\ref{genscal}) is present.  

A further indication that the standard picture of quantum criticality 
fails for the cuprates is found in the application of Eq. (\ref{dclimit})
to the universal scaling law of Homes, et. al.\cite{homes}.  Throughout the entire phase diagram of the cuprates, Homes, et. al.\cite{homes} have found
the empirical relationship,
\beq\label{homes}
\rho_s=\sigma_{\rm dc}(T_c^+)T_c
\eeq
between the superfluid density, $\rho_s$, the superconducting transition
temperature, $T_c$, and the dc conductivity just above $T_c$, $\sigma_{\rm dc}(T_c^+)$, holds within an accuracy of 5$\%$.  
By using the Drude formula for $\sigma_{\rm dc}$ and Tanner's\cite{tanner} empirical relationship between the superfluid and normal state densities, namely, $\rho_s=\rho_N/4$, Zaanen\cite{zaanen} has shown that Homes' Law reduces to Eq. (\ref{lint}). That is, the charge degrees of freedom in high $T_c$ superconductors are
at the quantum limit of dissipation, referred to by Zaanen as the Planckian limit.
Such Planckian dissipators are necessarily quantum critical.  However, the conclusion that Homes' Law represents a simple statement about the quantum limit of dissipation relies on the Drude formula, which, as we have discussed, 
has nothing to do with quantum criticality.  To assess the relevance of quantum
criticality to Homes' Law, it is more appropriate to use Eq. (\ref{dclimit}).
Substituting Eq. (\ref{dclimit}) into Eq. (\ref{homes}) results in a simple algebraic relationship\cite{kopp},
\beq\label{tcrs}
\rho_s\propto T_c^{(d-2)/z+1}
\eeq
 between the superfluid density and $T_c$.   Regardless of the exponent, this expression has a maximum whenever $T_c$ is maximized and hence is reminiscent of
the Uemura relationship\cite{uemura}, another empirical relationship valid only in the underdoped regime.  A key failure of the Uemura relationship is optimal doping where $\rho_s$ and $T_c$ are not simultaneously maximized.
Hence, we find that the form of the dc conductivity dictated by quantum criticality fails to capture the physics of Homes' Law, an empirical observation
valid regardless of doping.   Perhaps some as of yet to  be discovered
form of quantum criticality can explain Homes' law; but such an explanation must lie outside the one-parameter scaling hypothesis. 

The inability of Eq. (\ref{genscal}) to lead to a consistent account of
$T-$linear resistivity or Homes' Law\cite{homes} in the cuprates leaves us with three options. 1)
Either $T-$linear resistivity is not due to quantum criticality, 2)
additional non-critical degrees of freedom are necessarily the charge
carriers, or 3) perhaps some new theory of quantum criticality can be
constructed in which the single-correlation length hypothesis is relaxed. In
a scenario involving non-critical degrees of freedom, fermionic charge
carriers in the normal state of the cuprates could couple to a critical
bosonic mode. Such an account is similar to that in magnetic
systems\cite{qc1} in which fermions scatter off massless bosonic density or
spin fluctuations and lead to an array of algebraic forms for the
resistivity\cite{moriya} ranging from $T^{4/3}$ to $T^{3/2}$ in
antiferromagnetic and ferromagnetic systems, respectively.  While disorder
can alter the exponent\cite{rosch}, $T-$linear resistivity results only in a
restricted parameter space. Consequently, in the context of the cuprates, any
explanation of $T-$linear resistivity based on quantum criticality (as it is
currently formulated) must rely on the fortuitous presence of a bosonic mode
whose coupling to the fermions remains unchanged up to a temperature of
$T=1000K$.  Currently, no such mode which is strictly bosonic is known. This
is not surprising in light of the fact that numerous experimental systems
exist\cite{stewart} in which $T-$linear resistivity does not occur in the
quantum critical regime or $T-$ linear resistivity exists only at a single
point rather than a funnel-shaped region\cite{raffy,ando2}. These experiments imply
that the correspondence between quantum criticality and $T-$ linear
resistivity is not one of necessity.

What about new scenarios\cite{e1,e2} for quantum critical phenomena? For
example, an additional length scale, as is the case in deconfined quantum
criticality\cite{e2}, could provide the flexibility needed to obtain
$T-$linear resistivity while still maintining $z>0$.  A likely scenario is as
follows. Entertain the possibility that an additional length scale
$\tilde\xi$ is relevant which diverges as $\tilde\xi\propto \xi^a$, with
$a>1$. If in the calculation of the correlation volume entering
Eq.~(\ref{eq:lnZ}), one replaces $\xi^d$ with $\xi^d\to\ell^d=\xi^d\;
h(\tilde\xi/\xi)$, $h(y)=y^{-\lambda}$ a general scaling function, then one
is in essesence reducing the effective dimensionality such that $d\to
d^*=d-\lambda(a-1)$. $T-$linear resistivity results now if $z=2-d^*$.  The
reduction in the effecive dimensionality, $\lambda(a-1)$, can now be
fine-tuned so that $d^\ast\le 1$, thereby resulting in physically permissible
values of the dynamical exponent, $z\ge 1$. Nonetheless, such fine scripting
of two length scales is also without basis at this time.

Indeed, it is unclear what remedy is appropriate to square single parameter
scaling with $T-$linear resistivity in the cuprates.  It might turn out that
quantum criticality is not relevant to the problem. What is clear, however,
is that if $T-$linear resistivity is due to quantum criticality of the
degrees of freedom that carry the electrical charge, then a consistent theory
must be constructed to account for the breakdown of one-parameter scaling. In
fact, recent experiments on La$_2$CuO$_4$\cite{mag} find that the exponent of
the temperature prefactor of the magnetic suscpetibility\cite{mag} varies
across the critical region.  Perhaps this variation provides further evidence
that physics beyond the standard paradigms is necessary to explain the
magnetic and transport properties of the cuprates.

\acknowledgements
We thank Christos Panagopoulos for a key conversation which motivated this
work, A.~H. Castro Neto, E.~Fradkin, Y.~B.~Kim, W.~Klein, C.~Mudry, and
X.-G.~Wen for useful discussions.  We also thank the NSF Grants DMR-0305864
(P.~P.) and DMR-0305482 (C.~C.).  P.  Phillips also thanks the Aspen Center
for Physics for their hospitality during the writing of this paper.


\begin{thebibliography}{99}
\bibitem{finite}While it is certainly reasonable to assume that $\Sigma(0)$
  is finite at zero temperature, we cannot rule out the possibility that
  $\Sigma(0)$ might in fact diverge.  A possible scenario for how this state
  of affairs might obtain is that a dangerously irrelevant operator could
  control the conductivity.  In this case, $\Sigma(0)\sim 1/T^p$, and $T-$linear
  behaviour obtains if $p=(d-2)/z + 1$.  This can only occur above the upper
  critical dimension.  In this regime, all criticality is described by
  mean-field theory.  Hence, a possibility that cannot be eliminated at this
  time is that all criticality in the cuprates is inherently mean-field and
  dangerously irrelevant operators control the conductivity.  For strongly
  correlated systems, however, it is unlikely that criticality is inherently
  mean-field.
\bibitem{enscal}In systems with a Fermi surface, it is the Fermi energy that replaces the plasma frequency in Eq. (\ref{drude}). Transport in such systems is also incompatible with one-parameter scaling.
\end{thebibliography}
\end{document}